\documentclass[conference]{IEEEtran}
\IEEEoverridecommandlockouts

\makeatletter
\def\ps@IEEEtitlepagestyle{%
  \def\@oddfoot{\mycopyrightnotice}%
  \def\@oddhead{\hbox{}\@IEEEheaderstyle\leftmark\hfil\thepage}\relax
  \def\@evenhead{\@IEEEheaderstyle\thepage\hfil\leftmark\hbox{}}\relax
  \def\@evenfoot{}%
}
\def\mycopyrightnotice{%
  \begin{minipage}{\textwidth}
  \centering \scriptsize
  Copyright~\copyright~2024 IEEE. Personal use of this material is permitted. 
  Permission from IEEE must be obtained for all other uses,
  in any current or future media, including\\reprinting/republishing this material
  for advertising or promotional purposes, creating new collective works,
  for resale or redistribution to servers or lists, or reuse of any copyrighted
  component of this work in other works by sending a request to pubs-permissions@ieee.org.
  \end{minipage}
}
\makeatother

\usepackage{cite}
\usepackage{amsmath,amssymb,amsfonts}
\usepackage{graphicx}
\graphicspath{{./images/}}
\usepackage{textcomp}
\usepackage{xcolor}
\usepackage{multirow}
\usepackage{caption}
\usepackage{tabularx}
\usepackage{subfig}
\usepackage{mathtools}
\usepackage{color}
\usepackage{multicol}
\usepackage{listings} 
\usepackage{algorithm}
\usepackage{algpseudocode}
\usepackage{algorithmicx}

\usepackage{pifont}
\newcommand{\xmark}{\ding{55}}%

\usepackage{url}

\usepackage{breakurl}
\usepackage[breaklinks]{hyperref}
\usepackage{orcidlink}

\algdef{SE}[DOWHILE]{Do}{doWhile}{\algorithmicdo}[1]{\algorithmicwhile\ #1}

\def\BibTeX{{\rm B\kern-.05em{\sc i\kern-.025em b}\kern-.08em
    T\kern-.1667em\lower.7ex\hbox{E}\kern-.125emX}}
\begin{document}

\title{Hound: Locating Cryptographic Primitives in Desynchronized Side-Channel Traces Using Deep-Learning}

\author{
    \IEEEauthorblockN{Davide Galli\orcidlink{0009-0005-9430-7699}}
    \IEEEauthorblockA{\textit{DEIB} \\
    \textit{Politecnico di Milano}\\
    Milan, Italy \\
    davide.galli@polimi.it}
    \and
    \IEEEauthorblockN{Giuseppe Chiari\orcidlink{0009-0003-0659-1438}}
    \IEEEauthorblockA{\textit{DEIB} \\
    \textit{Politecnico di Milano}\\
    Milan, Italy \\
    giuseppe.chiari@polimi.it}
    \and
    \IEEEauthorblockN{Davide Zoni\orcidlink{0000-0002-9951-062X}}
    \IEEEauthorblockA{\textit{DEIB} \\
    \textit{Politecnico di Milano}\\
    Milan, Italy \\
    davide.zoni@polimi.it}
}

\maketitle

\begin{abstract}
Side-channel attacks allow the extraction of sensitive information from cryptographic
primitives by correlating the partially known computed data and the measured
side-channel signal.
Starting from the raw side-channel trace, the preprocessing of the side-channel
trace to pinpoint the time at which each cryptographic primitive is executed,
and, then, to re-align all the collected data to this specific time
represent a critical step to setup a successful side-channel attack. 
The use of hiding techniques has been widely adopted as a low-cost solution to
hinder the preprocessing of side-channel traces, thus limiting side-channel
attacks in real scenarios.
This work introduces Hound, a novel deep-learning-based pipeline to locate the
execution of cryptographic primitives within the side-channel trace even in
the presence of trace deformations introduced by the use of dynamic frequency
scaling actuators.
Hound has been validated through successful attacks on various cryptographic
primitives executed on an FPGA-based system-on-chip incorporating a RISC-V CPU
while dynamic frequency scaling is active. Experimental results demonstrate the possibility of identifying
the cryptographic primitives in DFS-deformed side-channel traces.   
\end{abstract}


\section{Introduction}
\label{sec:introduction}
\begin{figure*}[t]
	\centering
	\subfloat[Training Pipeline.]{\
		\includegraphics[scale=0.49]{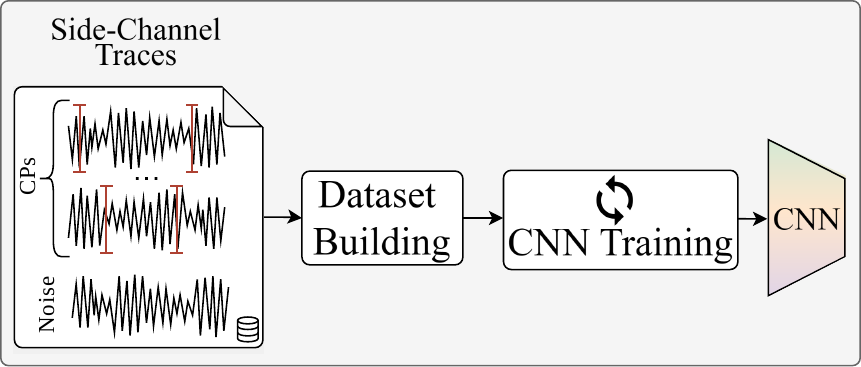}
		\label{sfig:pipeline_a}}
	\subfloat[Inference Pipeline.]{\
		\includegraphics[scale=0.49]{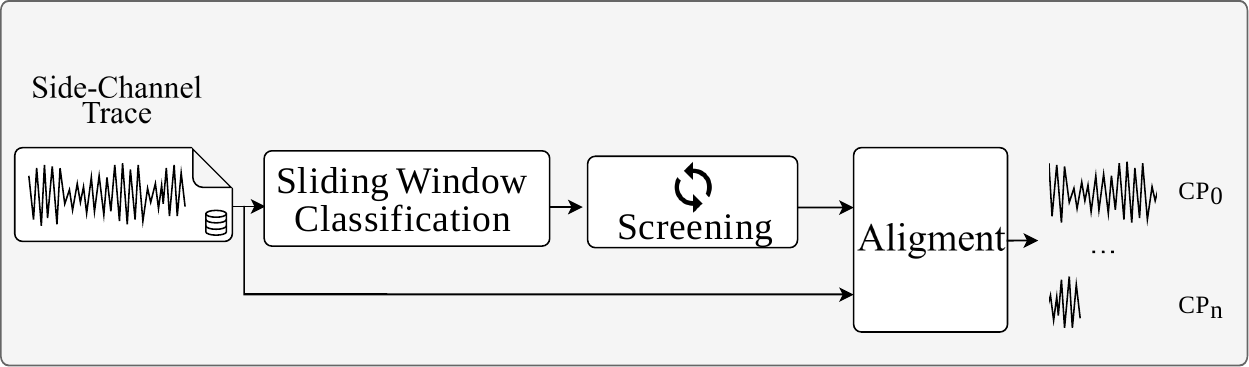}
		\label{sfig:pipeline_b}}
	\caption{Overview of the proposed Hound pipeline for locating cryptographic
			primitives in frequency-scaled side-channel traces, divided into \emph{training}
			and \emph{inference} pipelines.}
	\label{fig:pipeline}
\end{figure*}

Side-channel attacks pose a significant threat to modern cryptographic
implementations, even when the underlying algorithms boast mathematical soundness.
By exploiting vulnerabilities in the physical implementation of the
cryptographic primitives~(CPs), side-channel attacks can exploit the unintended
information leakage generated from electronic devices during the execution of
CPs~\cite{MOP08}. 
The leakage, often manifested in power consumption~\cite{KJJ99} or electromagnetic
emissions~\cite{AAR+03}, can be analyzed by attackers to potentially
reveal confidential information by correlating it with partially known processed data.

Traditionally, properly implemented masking~\cite{GP99, GMK16} and
hiding~\cite{CCD00, YWV+05, CK09} countermeasures offer robust protection
against side-channel attacks, leading to  successful security evaluations.
Masking countermeasures split sensitive intermediate values into independent shares
that are processed separately to reduce significantly the correlation
between each share and the secret key.
Hiding countermeasures introduce randomness into the targeted 
side-channel signal to degrade the signal-to-noise ratio.

Over the past two decades, researchers have proposed various methods
to enhance the effectiveness of side-channel analysis, such as differential
power analysis~(DPA)~\cite{KJJ99}, template attacks~\cite{CRR02},
correlation power analysis~(CPA)~\cite{BCO04}, as well as deep-learning-based
solutions~\cite{BPF+20, RWP+21}.
Notably, all the proposed techniques share two key requirements. First, they
require a large number of executions of the same CP with different inputs.
Second, the attacker needs to locate and align in time all the CP
executions in the side-channel trace to feed the attack method of choice.

In a controlled laboratory environment, the attacker has full access to the
target device, simplifying the security assessment of a CP implementation.
Security evaluation boards, like SASEBO SAKURA-II~\cite{sakura} and NewAE
CW305~\cite{cw305}, provide so-called trigger pins, facilitating the temporal 
alignment between CP executions and corresponding side-channel signals.

In contrast, attacking the implementation of a CP in the real world presents a
more significant challenge. Indeed, the attacker needs to locate the CP within
the side-channel trace. Specific settings still
permit a rough alignment of the measured side-channel signal with CP execution
even in real-world scenarios, e.g., by leveraging specific logic events
happening in the computing platform or by using pattern-matching techniques
applied to the side-channel trace to generate so-called virtual
triggers~\cite{newAeChipwhispererPRO,icWave}. 

However, the importance of pinpointing the executions of the CPs in the
side-channel trace even without a triggering infrastructure motivated several
contributions, i.e.,~\cite{BFP22, tches2021_semi_automatic_locating_sca}.
In a similar manner, the continuous improvement in the attack techniques
fueled the use of hiding techniques to hinder the task of locating a CP, thus
preventing the localization of the CPs in the side-channel trace as an
additional security countermeasure.

\cite{CGL+24} is the seminal work that presented a deep-learning
pipeline to locate the CPs within side-channel traces deformed by means of
random delay countermeasures. However, random delay introduces a limited trace
deformation, and its implementation requires custom hardware or software
implementations inducing non-negligible performance overheads~\cite{durvaux2012}.
In contrast, Dynamic Frequency Scaling~(DFS) actuators offer a cheap and standard
solution that has been widely investigated to implement hiding techniques. The
possibility of employing hundreds of operating frequencies allows to
effectively deform the side-channel trace at the point of preventing the correct
pinpointing of the CP and, consequently, invalidating the possibility of
attacking the target device even if the attacker can deploy an effective
side-channel attack.

\smallskip\noindent\textbf{Contributions -}
This paper introduces Hound, a deep-learning methodology for locating cryptographic
primitives within heavily randomized side-channel traces where the trace
deformation is obtained by means of a randomized DFS actuator.
Hound offers three contributions to the state of the art:
\begin{itemize}
	\item We propose a deep-learning pipeline that automatically locates and aligns
	CPs within side-channel power traces, even in the presence of DFS countermeasures.
	Thus, the need for triggering infrastructure is eliminated, which is a significant
    hurdle in real-world side-channel analysis.
	\item We evaluate the effectiveness of our solution across various cryptographic
	primitives. Furthermore, we perform successful side-channel attacks after CP
	localization and alignment to validate the quality of the results and compare
	our solution against state-of-the-art techniques.
	\item To facilitate further research and ensure reproducibility,
	we have released Hound as an open-source tool along with a dataset of relevant
	side-channel traces.
	The tool and dataset are available on GitHub~\footnote{
		\textcolor{magenta}{\underline{\url{https://github.com/hardware-fab/Hound}}}
	}.
\end{itemize}

The rest of the paper is organized into four sections. Section~\ref{sec:SoA}
discusses the academic and commercial tools to locate the CPs within 
side-channel traces. Section~\ref{sec:methodology} presents the proposed
Hound approach. Section~\ref{sec:expEval} details the experimental
results. Finally, Section~\ref{sec:conclusions} presents the conclusions.

\section{Related Works}
\label{sec:SoA}
Existing methods for locating CPs within side-channel
traces primarily rely on trigger signals from security evaluation boards
(SASEBO SAKURA-II~\cite{sakura} and NewAE CW305~\cite{cw305})
or pattern matching against pre-computed CP templates.
Commercially available FPGA-based devices, like the Riscure icWaves~\cite{icWave} and
NewAE ChipWhisperer Pro~\cite{newAeChipwhispererPRO}, offer virtual-triggering features,
producing a trigger pulse upon real-time identification of a pattern in the side-channel trace
being monitored.
Barenghi et al.~\cite{BFP22} introduced a method utilizing matched filters for
efficiently pinpointing the AES-128 cryptosystem within a power trace,
achieving robustness in environments with interrupts and busy waits that could change
the trace's profile. Nonetheless, this approach fails when it comes to recognizing
interrupted CPs, which are discarded by the matched filter. 
Becker et al.~\cite{cosade2016_becker} suggested a waveform-matching trigger system
designed to identify CPs by comparing them to a pre-calculated CP template.

All these methods need a pre-computed template of the 
CP to be localized to work correctly.
To address this issue, Trautmann et al. explored 
in~\cite{tches2021_semi_automatic_locating_sca}
a semi-automatic technique for locating CPs by building the template online, 
starting from some CP characteristics, such as the number of rounds.

\begin{figure*}[t]
	\centerline{\includegraphics[width=0.89\textwidth]{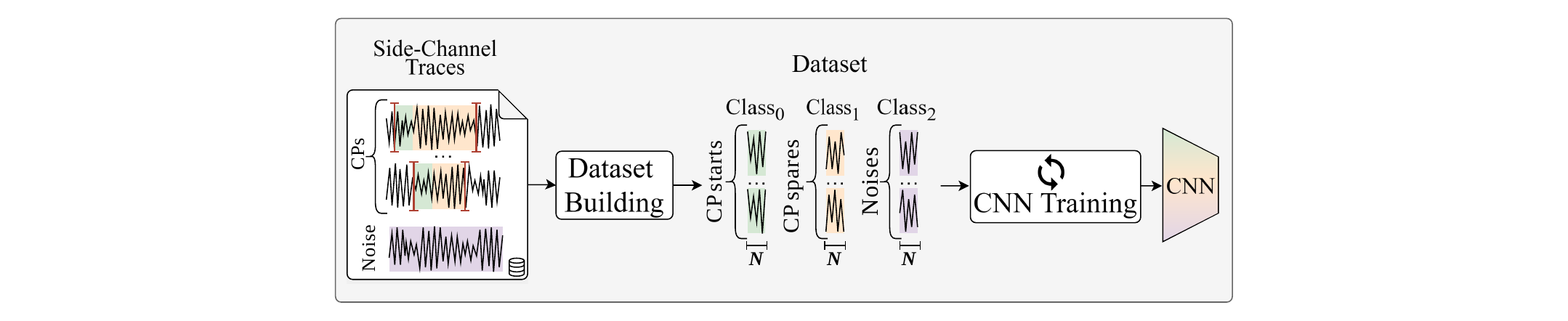}}
	\caption{Focus on the proposed Hound training pipeline, divided into \emph{Dataset Building} and
		\emph{CNN Training}. $Class_0$, $class_1$ and, $class_2$ contain 
		CP start part windows, CP spare parts windows, and noise windows, respectively.}
	\label{fig:train_pipeline}
\end{figure*}

However, using architectural-level techniques to morph the power trace
represents an easy-to-implement and effective countermeasure to deceive
pattern-matching-based solutions. 
For instance, employing time-sharing multithreading on a single-core
microcontroller~\cite{2016multiThreadCounterLocatingCO} can obstruct 
an attacker's ability to accurately locate CPs, with interrupt service 
routines further modifying the side-channel trace's form.
Consequently, more sophisticated methods for detecting CPs in the 
side-channel trace have been developed. Chiari et al.~\cite{CGL+24}
developed a deep-learning approach for identifying deformed CPs through random delay,
employing a Convolutional Neural Network~(CNN) trained 
to differentiate CPs from the remaining portions of the power trace.
Even though random delay-based hiding techniques generate 
some degree of CP deformation, they are limited in their ability 
to introduce significant variability in the power trace~\cite{durvaux2012}.

Despite these advancements, current state-of-the-art techniques 
struggle with highly deformed traces,
particularly when the computing platform employs effective 
hiding countermeasures like DFS~\cite{HDL+20}.
These countermeasures profoundly alter the timing of operations and 
continuously morph the side-channel signal,
making it challenging to isolate CPs using traditional methods.
To address this limitation, we propose a deep-learning-based technique 
for CP localization in power traces collected from platforms 
implementing DFS as a countermeasure.

\section{Methodology}
\label{sec:methodology}
\begin{figure*}[t]
	\centerline{\includegraphics[width=0.9\textwidth]{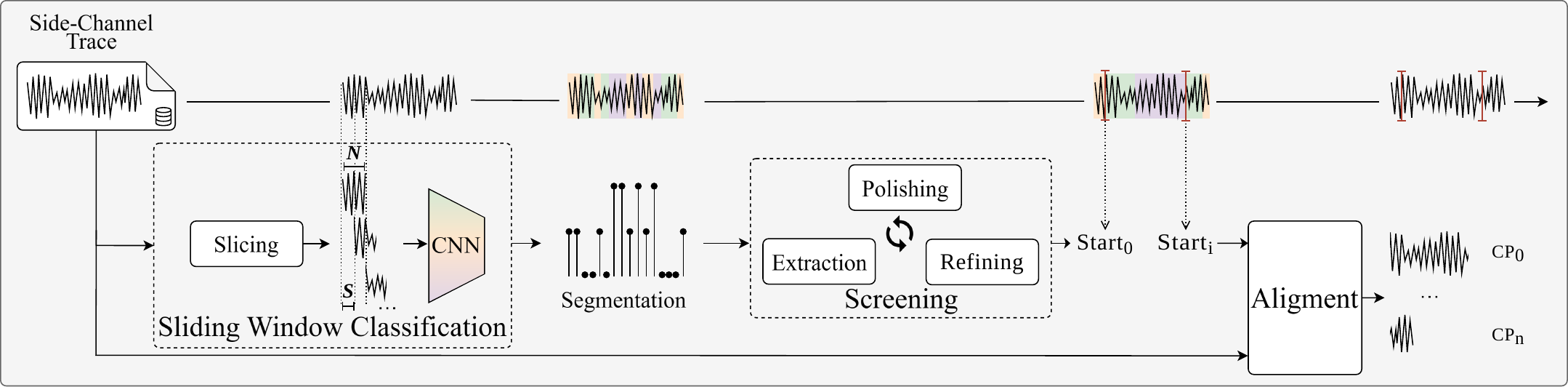}}
	\caption{Focus on the proposed Hound inference pipeline, divided into 
		\emph{Sliding Window Classification},
		\emph{Screening}, and \emph{Alignment}.}
	\label{fig:inf_pipeline}
\end{figure*}

This section introduces Hound, a novel deep-learning method for CP localization within 
highly deformed side-channel traces, specifically designed to address scenarios
where the target platform employs DFS as countermeasure to
effectively obfuscate the measurements.

\smallskip\noindent\textbf{Threat model -} 
Similarly to profiled attacks, we assume the attacker has access to
an identical copy of the target device, i.e., a clone.  However, we
focus on a realistic scenario where the attacker can only run selected
applications and monitor the resulting side channels without complete control
over the cloned device. Specifically, the attacker cannot activate
or deactivate the hiding countermeasure, namely DFS. Nonetheless, 
the attacker retains the ability to probe the clone device, enabling them to detect
when the CP initiates. This probing facilitates precise labeling of the beginning of
each CP on the clone device.

\smallskip\noindent\textbf{Hound -} 
Figure~\ref{fig:pipeline} depicts the proposed Hound methodology consisting of a
training and inference pipeline.
Figure~\ref{sfig:pipeline_a} reports the \emph{Training Pipeline}, which aims at creating
a classifier that can sort a window of the side-channel trace into three categories:
\emph{start of a CP},  \emph{spare part of a CP}, or just \emph{noise}~(see Figure~\ref{fig:train_pipeline}).
This framework locates where the CPs begin in the side-channel trace
without removing the obfuscation countermeasure.
We assume that the attacker leverages an exact copy of the target device to collect a noise
trace and some cipher traces for the training.
The noise trace is obtained by running general-purpose applications that are not the CP.
Such information is necessary to train a neural network to distinguish the beginning of
the CPs from the execution of other applications.
Each cipher trace is gathered while a single CP is running, where the attacker decides what
data and keys to use.
All measurements are collected with the obfuscation mechanism active since the attacker cannot
turn it off.
%
% three comments
% - hw CO, sw CO
%Notably, the proposed methodology works for software-executed and hardware-implemented CPs.
%In both scenarios, the noise trace contains the side channel obtained from the execution
%of other applications, i.e., applications different from the CP.
Starting with measured raw traces, the \emph{Dataset Building} stage creates a database 
of windows from the noise and cipher traces, each being $N$ samples long.
This database is then used to train a CNN classifier.

% inference pipeline
Figure~\ref{sfig:pipeline_b} reports the \emph{Inference Pipeline}. At inference
time, the goal is to identify the CPs within a novel side-channel trace from the
target device by employing the trained CNN.
%Utilizing the trained CNN, the \emph{Inference Pipeline}’s goal is to identify the CPs 
%within a novel side-channel trace from the target device.
%~(see \emph{Inference Pipeline} in Figure~\ref{fig:pipeline}).
This pipeline is divided into three key stages: \emph{Sliding Window Classification},
\emph{Screening}, and \emph{Alignment}.
The inference pipeline takes in a single side-channel trace and segments it to mark the
start of each CP.
Initially, the \emph{Slicing} stage processes a side-channel trace and generates a series of
windows to classify with the trained CNN.
Starting from the classified windows, representing a segmentation of the input trace,
the \emph{Screening} stage produces a sequence of time instants marking the start of each CP in
the trace.
Finally, the \emph{Alignment} stage chunks the original side-channel trace based on these
outputs, arranging the identified CPs accordingly.

The details related to the dataset creation and
the CNN architecture are discussed in
Section~\ref{ssec:train_meth}, while the sliding window
classification and the screening procedure are detailed in Section~\ref{ssec:inf_meth}.

%%%%%%%%%%%%%%%%%%%%%%%%%%%%%%%%%%%%%%%%%%%%%%%%%%%%%%%%%%%%%%%%%%%%%%%%%%%%%%%%%%%%
%%%%%%%%%%%%%%%%%%%%%%%%%%%%%%%%%%%%%%%%%%%%%%%%%%%%%%%%%%%%%%%%%%%%%%%%%%%%%%%%%%%%

\subsection{Training Pipeline}
\label{ssec:train_meth}
The Training Pipeline, detailed in Figure~\ref{fig:train_pipeline}, is divided into 
\emph{Dataset Building} and \emph{CNN Training} stages. Starting from a raw collection
of side-channel traces, the pipeline generates a dataset and trains a CNN classifier
to distinguish the starting of a CP.

\smallskip\noindent\textbf{Dataset Building -} 
The first stage in the training pipeline is \emph{Dataset Building}, which receives
a collection of side-channel traces as input and generates a dataset suitable for training the CNN.
%~(see \emph{Training Pipeline} in Figure~\ref{fig:pipeline}).
% types of traces
The attacker uses a clone of the target computing platform to create a noise
trace and a set of cipher traces. The noise trace captures the execution of 
various general-purpose applications, excluding the target CP.
Each cipher trace is collected while executing a single CP, where the attacker
controls the plaintext and secret key.
Notably, the attacker cannot disable the obfuscation mechanism, so the DFS remains active during
all trace collection.
% alternative to (virtual) trigger pins
Considering the proposed threat model, the attacker
can only execute a chosen application and measure the corresponding
side channel on the clone device.
However, the attacker can probe the clone device to detect the start of each CP.
In contrast, probing is not required for the noise trace.
Once trained, the CNN works on the target architecture that implements DFS
without the need for probing.
For each cipher trace $i$ of length $L_i$ samples, the starting $N$
samples are labeled as \textit{start of the CP}~(see $class_0$ in
Figure~\ref{fig:train_pipeline}).
The remaining $L_i - N$ samples are equally split into consecutive windows of
width $N$ and labeled as \textit{spare part of the CP}~(see $class_1$ in
Figure~\ref{fig:train_pipeline}).
Moreover, we extract a random set of $N$-sample windows from the noise trace
and we label each of them as \textit{noise}~(see $class_2$ in
Figure~\ref{fig:train_pipeline}).
Although the methodology aims to identify only the beginning of each CP,
the division into three classes allows CNN to maximize the distinction over the three cases.
%%%%%%%%%%%%%%%%%%%%%%%%%%%%%%%%%%%%%%%%%%%%%%%%%%%%%%%%%%%%%%%%%%%%%%%%%%%%%%%%%%%%
%%%%%%%%%%%%%%%%%%%%%%%%%%%%%%%%%%%%%%%%%%%%%%%%%%%%%%%%%%%%%%%%%%%%%%%%%%%%%%%%%%%%

% CNN architecture
\smallskip\noindent\textbf{Convolutional Neural Network -} 
The architecture of the 1D CNN is adapted from the CNN proposed in~\cite{CGL+24}.
The CNN takes a window of $N$ samples from the side-channel trace as input
and outputs a classification score vector.
The CNN’s structure is organized into six sequential components: a convolutional block,
a pair of residual blocks, a global average pooling layer, 
a fully-connected block, and a softmax layer.
Each convolutional block features a 1D convolutional layer, a batch normalization
layer~\cite{BN}, and a ReLU activation function.
The residual blocks~\cite{resnet}, each containing two convolutional blocks, are augmented with
shortcut connections to perform element-wise feature summation.
All the convolutional layers implement a kernel size of 64, a stride of 1,
and zero padding to maintain the sample count to $N$.
The first convolutional layer and the one in the first residual block implement
16 filters, while the subsequent residual block doubles the filter count to 32.
The global average pooling layer averages the obtained features over the
temporal dimension $N$.
The feature vector is then processed through two fully-connected layers with a ReLU
activation function.
The softmax layer completes the architecture, generating a $3$-dimensional 
vector with the classification scores for each class.

Notably, DFS introduces a significant amount of randomization in the side-channel trace.
To address this, the CNN proposed in~\cite{CGL+24} has been adapted to work with
three classes, i.e., \emph{start of the CP}, \emph{spare part of the CP}, and \emph{noise}.
This modification allows the CNN to learn the pattern of the highly obfuscated
side-channel trace better, maximizing the discrimination between the \emph{start of the CP} and
the other two classes.

%%%%%%%%%%%%%%%%%%%%%%%%%%%%%%%%%%%%%%%%%%%%%%%%%%%%%%%%%%%%%%%%%%%%%%%%%%%%%%%%%%
%%%%%%%%%%%%%%%%%%%%%%%%%%%%%%%%%%%%%%%%%%%%%%%%%%%%%%%%%%%%%%%%%%%%%%%%%%%%%%%%%%
\begin{algorithm}[t]
	\caption{Screening}
	\label{alg:screening}
	\hspace*{\algorithmicindent} \textbf{Input}: $seg$, $s$, $k$, $avgCP$\\
	\hspace*{\algorithmicindent} \textbf{Output}: $starts$
	\begin{algorithmic}[1]
		\State $starts \gets \texttt{[]}$
		\Do
		\State $polishedSeg \gets \texttt{POLISH(}seg, k\texttt{)}$
		\State $newStarts \gets \texttt{EXTRACT(}polishedSeg, s\texttt{)}$
		\State $k, seg \gets \texttt{REFINE(}k, avgCP, newStarts, seg\texttt{)}$
		\State $starts\texttt{.update(}newStarts \texttt{)}$
		\doWhile{$seg$ not empty $\&\&$ $k \ge 1$}
		\State \Return $starts$
		\\
		\Procedure{polish}{$seg$, $k$}
		\State $polishedSeg \gets \texttt{majorityFilter(}seg, k\texttt{)}$
		\State \Return $polishedSeg$
		\EndProcedure
		\Procedure{extract}{$seg$, $s$}
		\State $newStarts \gets$ [  ]
		\For{$i, sample$ in \texttt{enumerate($seg$)}}
		\If{$sample$ is falling edge to class 0}
		\State $newStarts\texttt{.update(}i \times s \texttt{)}$
		\EndIf
		\EndFor
		\State \Return $newStarts$
		\EndProcedure
		\Procedure{refine}{$k$, $avgCP$, $starts$, $seg$}
		\State $k \gets \texttt{refineK(}k \texttt{)}$
		\State $minCP \gets \texttt{refineMin(}avgCP, starts\texttt{)}$
		\State $refineSegs \gets \texttt{[]}$
		\State $\text{// Look if can be still two consecutive CPs}$
		\For{$s, s_{next}$ in $starts$}
		\If{$s_{next} - s > 2 \times minCP$}
		\State $subSeg \gets seg[s:s_{next}]$
		\State $refineSegs\texttt{.update(}subSeg \texttt{)}$
		\EndIf
		\EndFor
		\State \Return $k, refineSegs$
		\EndProcedure
	\end{algorithmic}
	
\end{algorithm}

\subsection{Inference Pipeline}
\label{ssec:inf_meth}
The Inference Pipeline, detailed in Figure~\ref{fig:inf_pipeline}, is divided into
\emph{Sliding Window Classification}, \emph{Screening}, and \emph{Alignment} stages.
Leveraging the trained CNN classifier, the pipeline locates the starting sample of each CP
present in a new side-channel trace. 
\smallskip\noindent\textbf{Sliding Window Classification -} 
The first stage in the inference pipeline is \emph{Sliding Window
Classification}. It takes a new side-channel trace, slices it into $N$-sample
windows, and uses the trained CNN to output a classification score to label
each window as \emph{start of the CP}, \emph{spare part of the CP}, or
\emph{noise}~(see \emph{Sliding Window Classification} in Figure~\ref{fig:inf_pipeline}).
The \emph{Slicing} stage implements a sliding window procedure to slice the
side-channel trace in input. It takes three inputs, i.e., the
side-channel trace, the size of the sliding window~($N$),
and the stride~($s$), and outputs an ordered set of $N$-sample windows to feed the
CNN.

The CNN's softmax output is a $3$-dimensional probability distribution of classes,
classifying each $N$-sample window as \emph{start of the CP},
\emph{spare part of the CP} or \emph{noise}.
The inference process considers the class with the highest probability outcome among
the three as the CNN's output.

Since the CNN's output might be noisy and not always suitable for directly determining 
the exact location of the CPs, the inference pipeline incorporates a subsequent 
\emph{Screening} stage to refine the CNN's output (see \emph{Screening} in Figure~\ref{fig:inf_pipeline}).

%%%%%%%%%%%%%%%%%%%%%%%%%%%%%%%%%%%%%%%%%%%%%%%%%%%%%%%%%%%%%%%%%%%%%%%%%%%%%%%%%%
%%%%%%%%%%%%%%%%%%%%%%%%%%%%%%%%%%%%%%%%%%%%%%%%%%%%%%%%%%%%%%%%%%%%%%%%%%%%%%%%%%

\smallskip\noindent\textbf{Screening -} 
The \emph{Screening} stage refines the segmentation output from the 
\emph{Sliding Window Classification} stage to identify the beginning of the CPs in the side-channel
trace.
Algorithm~\ref{alg:screening} highlights the screening procedure.
The screening algorithm~(see Algorithm~\ref{alg:screening}) takes
the segmentation output from the \emph{Sliding Window Classification} stage,
denoted as $seg$, and some parameters, i.e., the stride $s$ used for the sliding window,
the kernel size $k$ for the polish procedure (see line 10 of Algorithm~\ref{alg:screening}),
and the average CP length $avgCP$. In the end, it returns the list $starts$ 
of samples corresponding to the beginning of each CP instance 
in the processed side-channel trace.

The algorithm is an iterative process that involves three main steps:
\texttt{POLISH}, \texttt{EXTRACT}, and \texttt{REFINE}.
The core idea behind Algorithm~\ref{alg:screening} is to start with an aggressive polish and
incrementally refine it by attempting to uncover new CPs.
The polishing step (see line 10 of Algorithm~\ref{alg:screening}) employs a
majority filter across the whole segmentation.
This filter traverses the input signal for a given kernel size $k$, replacing each point
with the most frequently occurring value within a $k$-sized window.
The extraction step (see line 14 of Algorithm~\ref{alg:screening}) identifies the
initial sample of each CP. It pinpoints the indices of the falling edge down to class 0,
meaning those samples with a value of 0 where the preceding value is not 0.
These samples are then scaled by $s$, i.e., the stride in the sliding window classification.
Finally, the refining step (see line 23 of Algorithm~\ref{alg:screening}) fine-tunes
the algorithm’s parameters by reducing kernel size $k$ and generating a list of 
sub-segmentation for subsequent iterations. These sub-segments encompass those regions
of the input segmentation for which multiple hidden CPs might still exist. 
The process begins by determining the minimum length of a CP $minCP$
(see line 25 of Algorithm~\ref{alg:screening}),
which is derived by comparing the input's average CP length and
the CP's starting points identified so far.
Next, the algorithm examines pairs of consecutive CP starting points
(see line 28 of Algorithm~\ref{alg:screening}).
If the distance between them exceeds twice the $minCP$,
it indicates the potential presence of additional CPs within that interval.
Consequently, the corresponding sub-segmentation is queued for further processing.

The algorithm persists until no further refinement is possible, i.e., when $k$
drops below one or the refining step yields an empty list of sub-segmentations $seg$
(see line 7 of Algorithm~\ref{alg:screening}).

%%%%%%%%%%%%%%%%%%%%%%%%%%%%%%%%%%%%%%%%%%%%%%%%%%%%%%%%%%%%%%%%%%%%%%%%%%%%%%%%%%
%%%%%%%%%%%%%%%%%%%%%%%%%%%%%%%%%%%%%%%%%%%%%%%%%%%%%%%%%%%%%%%%%%%%%%%%%%%%%%%%%%

\smallskip\noindent\textbf{Alignment -}
The last stage of the inference pipeline is \emph{Alignment}.
It leverages the CP starting points identified by the previous stages to chunk
the input side-channel trace into temporally aligned segment.

\section{Experimental Evaluation}
\label{sec:expEval}
\begin{figure*}[t]
    \centering
    \subfloat[AES.]{\
		\includegraphics[width=0.18\textwidth]{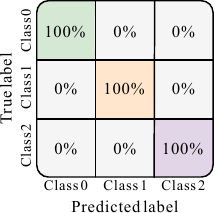}
    \label{sfig:cm_aes}}
    \qquad
    \subfloat[AES Masked.]{\
		\includegraphics[width=0.18\textwidth]{CM_100.pdf}
    \label{sfig:cm_aesMaksed}}
    \qquad
    \subfloat[Clefia.]{\
		\includegraphics[width=0.18\textwidth]{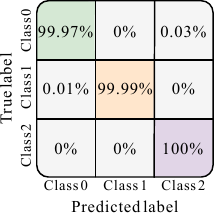}
    \label{sfig:cm_clefia}}
    \qquad
    \subfloat[Camellia.]{\
		\includegraphics[width=0.18\textwidth]{CM_100.pdf}
    \label{sfig:cm_camellia}}
    \caption{Test confusion matrices for the different cryptosystems affected by DFS.}
    \label{fig:aes_confusion_matrix_test}
\end{figure*}

\begin{table*}[t]
	\centering
	\caption{Parameters for each pipeline stage and metrics score over all the tested ciphers.}
  \setlength{\extrarowheight}{0.11em}  
  \begin{tabular}{|c|c||c|c|c|c||c|c|c||c|c|c|}
    	\hline
        \multirow{2}{*}{\textbf{Cipher}}     & \textbf{General-Purpose}       & \multicolumn{4}{c||}{\textbf{Inference parameters}} & \multicolumn{3}{c||}{\textbf{CNN hyperparameters}} &\multicolumn{3}{c|}{\textbf{Metrics}}  \\ \cline{3-12}
                                             & \textbf{Applications}& \textbf{\emph{avgCP}} & \textbf{\emph{k}} & \textbf{\emph{N}} & $\textbf{s}$ & \textbf{batch size} & \textbf{lr} & \textbf{dropout} & \textbf{Hits} & \textbf{Mean IoU} & \textbf{Std IoU} \\ \hline\hline
        \multirow{2}{*}{\textbf{AES}}	       & \checkmark & 145k & 150 & 10k   & 62 & \multirow{2}{*}{256} & \multirow{2}{*}{0.01} &\multirow{2}{*}{0.2}  & 100\% & 97.01\% & 1.65\% \\ \cline{2-6}\cline{10-12}
                                             & \xmark     & 120k & 150 & 10k   & 62 &  &  &  & 100\% & 93.62\% & 2.75\% \\ \hline 
        \multirow{2}{*}{\textbf{AES Masked}} & \checkmark & 50k  & 10  & 5k    & 50 &  \multirow{2}{*}{256} &  \multirow{2}{*}{0.007} & \multirow{2}{*}{0.35} & 100\% & 97.13\% & 1.85\% \\ \cline{2-6}\cline{10-12}
                                             & \xmark     & 50k  & 150 & 5k    & 50 & & & & 100\% & 95.05\% & 1.43\% \\ \hline
        \multirow{2}{*}{\textbf{Clefia}} 	   & \checkmark & 80k  & 150 & 3k    & 80 & \multirow{2}{*}{256}  & \multirow{2}{*}{0.007}  & \multirow{2}{*}{0.3} & 100\% & 97.90\% & 2.88\% \\ \cline{2-6}\cline{10-12}
                                             & \xmark     & 80k  & 150 & 3k    & 80 & & & & 100\% & 98.46\% & 0.76\% \\ \hline 
        \multirow{2}{*}{\textbf{Camellia}} 	 & \checkmark & 4.4k & 80	 & 1.1k  & 50 & \multirow{2}{*}{128} &  \multirow{2}{*}{0.007} & \multirow{2}{*}{0.4} & 100\% & 91.92\% & 4.79\%  \\ \cline{2-6}\cline{10-12}
                                             & \xmark     & 4.1k & 63	 & 1.1k  & 50 & & & & 100\% & 93.09\% & 3.63\%  \\ \hline 
    \end{tabular}
	\label{tbl:pipeline_params}
\end{table*}

This section presents the experimental results of Hound in four parts.
Section~\ref{ssec:exp_setup} outlines the experimental setup.
Experimental results targeting the CNN training and inference pipelines are
discussed in Section~\ref{ssec:train_exp} and Section~\ref{ssec:inf_exp},
respectively.
A comprehensive example workflow, including a successful side-channel attack 
and a comparison with state-of-the-art proposals,
is provided in Section~\ref{ssec:sca_exp}.

%%%%%%%%%%%%%%%%%%%%%%%%%%%%%%%%%%%%%%%%%%%%%%%%%%%%%%%%%%%%%%%%%%%%%%%%%%%%%%%%%%
%%%%%%%%%%%%%%%%%%%%%%%%%%%%%%%%%%%%%%%%%%%%%%%%%%%%%%%%%%%%%%%%%%%%%%%%%%%%%%%%%%

\subsection{Experimental Setup: Hardware and Software}
\label{ssec:exp_setup}

We utilized the NewAE CW305 development board~\cite{cw305} as our validation platform,
designed specifically to facilitate high-precision, low-noise side-channel measurements.
This board incorporates an AMD Artix7-100 FPGA.
Power traces were captured using a Picoscope 5244d digital sampling oscilloscope~(DSO),
operating at a sampling rate of 125~Msamples/s and a resolution of 12~bits.
For our reference computing platform, we employed a 32-bit RISC-V System-on-Chip~\cite{DGC+24}
implemented on the FPGA. Taking steps from~\cite{GGF+24}, we implemented the Dynamic
Frequency Scaling mechanism coupled with a true random number generator~(TRNG)~\cite{GGF+22}
to deliver a novel actuator that can randomize the operating frequency, thus
introducing non-negligible deformations in the collected side-channel traces.
The TRNG generates a random number used to select a frequency from a pool
of 760 available frequencies ranging from 5~MHz to 100~MHz, with a step of 125~kHz.
Notably, DFS remained enabled throughout the entire experimental phase.
Consequently, a new frequency is requested as soon as a configuration is locked. 
As a result, each cryptographic primitive experiences a variable number of clock
frequency reconfigurations, e.g., an average of 41 reconfigurations per AES encryption,
leading to significant trace deformation.
As the cryptographic primitives of choice, we selected the constant-time,
unprotected version of three ciphers, i.e., AES-128, Clefia-128, and Camellia-128,
from the OpenSSL software codebase~\cite{openSSL}, and a masked
version of Tiny-AES-128~\cite{tinyAES}.

%%%%%%%%%%%%%%%%%%%%%%%%%%%%%%%%%%%%%%%%%%%%%%%%%%%%%%%%%%%%%%%%%%%%%%%%%%%%%%%%%%
%%%%%%%%%%%%%%%%%%%%%%%%%%%%%%%%%%%%%%%%%%%%%%%%%%%%%%%%%%%%%%%%%%%%%%%%%%%%%%%%%%

\subsection{Training Evaluation}
\label{ssec:train_exp}
This section details the evaluation of the performance of the CNN 
by elaborating on the dataset-building process and the training metrics.

The training of the CNN leverages a \emph{NVIDIA 1080 Ti} employing the PyTorch
software programming framework.
The training datasets were built following the methodology outlined in
Section~\ref{ssec:train_meth}, starting from a collection of $262\,144$ 
power traces. The three classes, namely \emph{start of a CP},
\emph{spare part of a CP}, and \emph{noise},
were balanced with equal representation of 33\% components each.
Table~\ref{tbl:pipeline_params} reports the window sizes $N$ for each cipher.
As a general guideline, the window size $N$ is adjusted to approximate one 
round of the targeted CP. For fast encryption algorithms, like Camellia,
the window size is increased to capture a more significant portion of the CP without compromising
the quality of results.
The windows corresponding to the ciphers, i.e., 
\emph{start of a CP} and \emph{spare part of a CP},
are evenly distributed across the key bytes. 
Following standard deep-learning practices, the collected datasets were divided 
into training, validation, and testing sets, constituting 80\%, 10\%, and 10\% 
of the total, respectively.

Each neural network underwent training for $25$ epochs, 
where an epoch encompasses a complete iteration over the training set. 
The Adam optimizer~\cite{Adam} was employed to minimize the cross-entropy loss,
paired with the one-cycle learning rate scheduler.
The validation error was assessed after each epoch, and the network with the lowest 
error was chosen for further evaluation. Subsequently, this network was employed to 
assess the performance of unseen traces during the inference phase.
Since DFS introduces significant trace deformation with high variability,
we noticed how CNNs that did not reach an accuracy of at least 99\% were not always 
able to classify the CPs in the sliding windows procedure correctly.
Hyperparameters such as training batch size, learning rate~(lr), and dropout were
carefully chosen to address this issue~(see the \emph{CNN hyperparameters} column in
Table~\ref{tbl:pipeline_params}).

A distinct CNN model has been trained using a custom dataset for each cipher under examination. 
As an indicator of the effectiveness of the trained CNNs, their confusion matrices are
presented in Figure~\ref{fig:aes_confusion_matrix_test}.
The column indices represent the true classes, while the row indices represent
the predicted ones.
Notably, the trained classifiers can discriminate excellently between
the three classes, as highlighted by the high percentages on the main diagonal of each matrix.

\subsection{Inference Evaluation}
\label{ssec:inf_exp}
A brief experimental campaign was conducted to evaluate the best pipeline parameters.
Table~\ref{tbl:pipeline_params} displays the values of the average cipher
length $avgCP$, the strides $s$, and the initial kernel size $k$ for each cipher.
We evaluate the performance of the \emph{Sliding Window Classification} and \emph{Screening}
stages by computing two quality metrics: \textit{hits} and \textit{Intersection over Union}~(IoU).
\textit{Hits} represent the proportion of correctly identified CPs relative to the total number
of actual CPs within the trace.
The \textit{IoU}, on the other hand, measures the normalized overlap between the predicted CP
starting point and the ground truth, taking into account the CP's overall length.
A higher \textit{IoU} score signifies better agreement between the predicted and actual CP starting points.
It is formally defined in Equation~\ref{eq:iou}, where $P_{i}$ and $GT_{i}$ are the segments 
corresponding to the CP $i$ in the predicted and ground-truth segmentation, respectively.
\begin{equation}
\label{eq:iou}
IoU_i = \frac{|P_i \cap GT_i|}{|P_i \cup GT_i|}
\end{equation}
Since Hound focuses on pinpointing the CP's beginning within the trace,
the \textit{IoU} metric disregards the CP's ending point.
The predicted ending point is always assumed to be accurate.
Table~\ref{tbl:pipeline_params} shows the mean and standard deviation of \textit{IoU}, 
where a higher score corresponds to a better segmentation.

We conducted tests on the inference pipeline for each cipher using consecutive cipher
executions and encryptions interleaved with random general-purpose applications.
The segmentation \textit{hits} score achieved 100\% for every cryptographic algorithm 
in both scenarios, i.e., consecutive encryption and interleaved with 
noisy general-purpose applications,
consistently managing to identify all executions.
The \textit{IoU} score, on the other hand, ranges from 91.92\% to 98.46\% for the different ciphers.
The results demonstrate the robustness of the proposed methodology in locating CPs with high
accuracy and precision, even in the presence of DFS-induced trace deformation.

\begin{figure}[t]
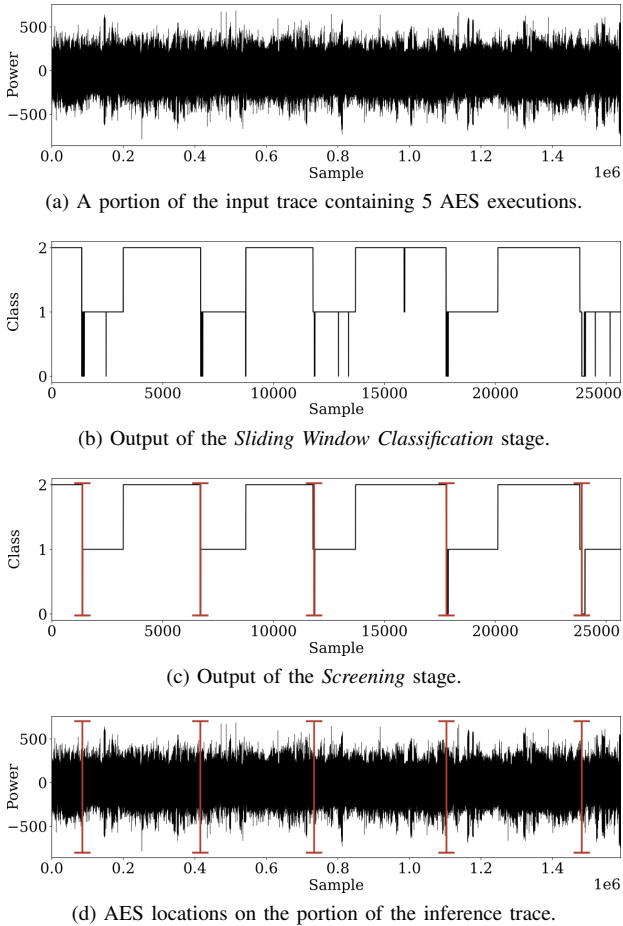

    \centering
    \subfloat[A portion of the input trace containing 5 AES executions.]{\
		\includegraphics[width=0.45\textwidth]{power_trace.pdf}
    \label{sfig:power}}\\
    \subfloat[Output of the \textit{Sliding Window Classification} stage.]{\
		\includegraphics[width=0.45\textwidth]{swc.pdf}
    \label{sfig:swc}}\\
    \subfloat[Output of the \textit{Screening} stage.]{\
		\includegraphics[width=0.45\textwidth]{screening.pdf}
    \label{sfig:screeining}}\\
    \subfloat[AES locations on the portion of the inference trace.]{\
		\includegraphics[width=0.45\textwidth]{aligned.pdf}
    \label{sfig:aligned}}
    \caption{Example of Hound inference pipeline applied to a
    side-channel trace affected by DFS that contains 5 AES
    executions mixed with general-purpose applications.}
    \label{fig:pipeline_example}
\end{figure}

\begin{table}[t]
	\centering
	\caption{Segmentation and side-channel attack~(SCA) results targeting AES-128.
    Reported results consider the presence (or not) of general-purpose applications
    interleaved with the CPs in the side-channel trace.
	}
  \setlength{\extrarowheight}{0.11em}
		\begin{tabular}{|c|c|c|c|c|}
    \hline
                                            & \textbf{Hiding}        & \textbf{General-Purpose} 	     & \textbf{Hits}   & \textbf{SCA} \\
                                            & \textbf{Mechanism}     & \textbf{Applications} & (\%)            & (N. CPs)  \\ \hline\hline
    \multirow{2}{*}{\textbf{\cite{BFP22}}}  & \multirow{2}{*}{None}  & \checkmark	           & 0\%             & \xmark  	\\ \cline{3-5}		
			                                      &				                 & \xmark                & 0\%	           & \xmark   \\ \hline
    \multirow{2}{*}{\textbf{\cite{tches2021_semi_automatic_locating_sca}}} &\multirow{2}{*}{None}& \checkmark&0\%& \xmark \\ \cline{3-5}		
                                            &                       & \xmark                 & 0\%             &	\xmark  \\ \hline
    \multirow{2}{*}{\textbf{\cite{CGL+24}}} & Random Delay          & \checkmark             & 1637\%          & 	\xmark  \\ \cline{3-5}		
	                                          & Insertion             & \xmark                 & 879\%           &	\xmark  \\ \hline         
    \multirow{2}{*}{\textbf{Hound}}         & \multirow{2}{*}{DFS}  & \checkmark             & \textbf{100\%}  & 	571     \\ \cline{3-5}		
                                            &                       & \xmark                 & \textbf{100\%}  &	74      \\ \hline
        \end{tabular}
	\label{tbl:AES_results}
\end{table}

%%%%%%%%%%%%%%%%%%%%%%%%%%%%%%%%%%%%%%%%%%%%%%%%%%%%%%%%%%%%%%%%%%%%%%%%%%%%%%%%%%
%%%%%%%%%%%%%%%%%%%%%%%%%%%%%%%%%%%%%%%%%%%%%%%%%%%%%%%%%%%%%%%%%%%%%%%%%%%%%%%%%%

\subsection{The Complete Attack Flow}
\label{ssec:sca_exp}
This section showcases the efficacy of Hound in locating
cryptographic primitives obfuscated by DFS.
We present a comprehensive attack flow that takes an unknown side-channel trace
and retrieves the secret key. The key is retrieved using a deep-learning-based
attack as an effective side-channel analysis technique.

Figure~\ref{fig:pipeline_example} shows the results of applying each
step of the proposed inference pipeline to locate the CPs into a portion of a
trace containing five executions of AES-128, i.e., the CP. The entire
side-channel trace contains 367 CPs collected from a platform
implementing a randomized dynamic frequency scaling mechanism.
Starting from the portion of the side-channel trace in input,
Figure~\ref{sfig:swc} depicts the segmentation output of the \emph{Sliding
Window Classification} stage that correctly highlights the presence of five
CPs interleaved with random general-purpose applications.
Notably, it is easy to detect also the end of the AES encryptions by
looking at \emph{Class 1}, as the classifier is trained to recognize the
spare part of the CPs.
The number of samples in the x-axis of the plot is reduced by a factor of $62$,
equal to the value of stride $s$ during the classification.
The \emph{Screening} stage cleans the segmentation output by
removing spurious activations of the various classes~(see
Figure~\ref{sfig:screeining}). The time instants locating the beginning of
the five CPs are applied to the original side-channel trace, thus allowing an
easy time realignement~(see Figure~\ref{sfig:aligned}).

As an effective side-channel attack, we employed a CNN developed using the tool
proposed by~\cite{RWP+21}.
The attacking CNN targets the \emph{sub-byte} intermediate and has been
trained starting from the same side-channel collection described in
Section~\ref{ssec:train_meth}. A minor sub-sampling is used to ease the training
process of the attacking CNN, reducing the input size and expediting the process.
Column SCA in Table~\ref{tbl:AES_results} enumerates the number of CPs
needed to reach a guessing entropy of 1, i.e., guess the secret key correctly.

\smallskip\noindent\textbf{Comparison against state of the art -}
Building on the prior evaluation, Table~\ref{tbl:AES_results} further stresses
the effectiveness of the proposed methodology,
detailing the attack results under various scenarios.
The target CP is AES-128. We compare our approach against
three state-of-the-art proposals, i.e., \cite{BFP22}, 
\cite{tches2021_semi_automatic_locating_sca}, and \cite{CGL+24}.
For each state-of-the-art approach, we report the hiding mechanism for
which the methodologies were designed by their authors~(see the \emph{Hiding
Mechanism} column in Table~\ref{tbl:AES_results}).
We considered the random CPs' execution within a set of noisy
general-purpose applications and their execution without any noisy
application~(see the \emph{General-Purpose Applications}
column in Table~\ref{tbl:AES_results}).
For each analyzed scenario, Hound can correctly identify the
beginning of all the CPs in the side-channel trace, leading to a successful
side-channel attack.
In contrast, the three state-of-the-art methodologies fail to locate the
CPs in the side-channel trace due to the high DFS obfuscation, and thus,
the subsequent side-channel attack is unsuccessful.
More specifically, \cite{BFP22} and \cite{tches2021_semi_automatic_locating_sca}
achieve a 0\% hit rate, indicating they cannot locate any CPs in the trace.
On the other hand, \cite{CGL+24} exhibits anomalously high hit rates of 879\% and
1637\% for the two scenarios. These excessively high rate values highlight a
significant number of false positives determining the subsequent failure of the
side-channel attacks.

\section{Conclusions}
\label{sec:conclusions}
We presented Hound, a novel deep-learning technique for pinpointing
cryptographic primitives within highly desynchronized side-channel traces.
Unlike existing methods, Hound successfully locates the CPs in side-channel
traces collected from computing platforms implementing dynamic frequency scaling
mechanism as an effective hiding countermeasure.

We conducted a comprehensive evaluation to locate both sequences of
consecutive cryptographic primitives and those interleaved with other applications.
The evaluation considered different cryptographic primitives.
The experimental results, obtained by running a diverse set of applications on
an FPGA-based RISC-V processor, validate the effectiveness of our approach in
enabling successful side-channel attacks.
These results also emphasize the shortcomings of current state-of-the-art solutions.

To promote reproducibility and future research, we have made Hound publicly
available as open-source software, along with a collection of test traces.

\bibliographystyle{IEEEtran}
\bibliography{bibliography}

\end{document}